# Are Heating Events in the Quiet Solar Corona Small Flares? – Multiwavelength Observations of Individual Events

Säm Krucker[1,2] and Arnold O. Benz[1]
[1]*Institute of Astronomy, ETH-Zentrum, CH-8092 Zurich, Switzerland*
[2]*Space Sciences Laboratory, University of California, Berkeley CA 94720, USA*

**Abstract.** Temporary enhancements of the coronal emission measure in a quiet region have been shown to constitute a significant energy input. Here some relatively large events are studied for simultaneous brightenings in transition region lines and in radio emission. Associated emissions are discussed and tested for characteristics known from full-sized impulsive flares in active regions. Heating events and flares are found to have many properties in common, including *(i)* associated polarized radio emission, which usually precedes the emission measure peak (Neupert effect) and sometimes has a non-thermal spectrum, and *(ii)* associated and often preceding peaks in O V and He I emission. On the other hand, heating events also differ from impulsive flares: *(i)* In half of the cases, their radio emission at centimeter waves shows a spectrum consistent with thermal radiation, *(ii)* the ratio of the gyro-synchrotron emission to the estimated thermal soft X-ray emission is smaller than in flares, and *(iii)* the associated emission in the O V transition region line shows red shifts and blue shifts, indicating upflows in the rise phase and downflows in the decay phase, respectively. Nevertheless, the differences seem to be mainly quantitative, and the analyzed heating events with thermal energies around $10^{26}$erg may in principle be considered as microflares or large nanoflares, thus small versions of regular flares.

## 1. Introduction

Impulsive enhancements of the emission measure are a major energy input in quiet regions of the solar corona, and thus contribute substantially to the heating (Benz & Krucker 1998; Krucker & Benz 1998). They cause an incessant flickering of the coronal brightness in soft X-rays and high-temperature EUV lines. What causes these impulsive additions to the coronal plasma? Are they just miniature flares?

In the *transition region* of the quiet Sun, ubiquitous brightness changes are well known from EUV lines (Brueckner & Bartoe 1983). Pulsating sources, called blinkers, have been detected (Harrison 1997) and bipolar jets have been reported and attributed to reconnection events in the chromosphere (Innes et al. 1997).

In the quiet *corona*, X-ray bright points have been analyzed simultaneously in several wavelengths. Typical thermal energies of X-ray bright points range from $10^{27} - 10^{29}$erg (Golub et al. 1974). They are relatively infrequent, last for several hours and may be the extension of the heating event distribution to large energies. Habbal et al. (1986), Fu et al. (1987) and others have found associated radio emission in all cases. Kankelborg et al. (1997) observed associated Ly-$\alpha$ emission at the footpoints of coronal bright points and conclude that the brightening is consistent with classical heat conduction from the corona.

Impulsive enhancements of the emission measure are many orders of magnitude more frequent and less energetic than X-ray bright points. Benz & Krucker (1999) have analyzed emissions associated to heating events by cross-correlating common pixels in time. The energies sampled ranged from $10^{24}$erg to a few times $10^{26}$erg per event, and the temperature reached up to $1.5\times10^{6}$K. They found that the peak correlation of time profiles at different wavelengths is shifted in time. The 3.6 cm and 6 cm radio emissions lead the Fe XII emission by about one minute, and the He I and O V line emissions precede Fe XII by about 5 minutes. In all cases, the peak value of the cross-correlation was found to be significant, but very low. There can be at least two reasons for the low peak value:

1. Coronal brightenings are spatially displaced by often more than one pixel (2000 km) from transition region brightenings (Benz et al. 1997). The spatial cross-correlation averages out the individual displacements.

2. The coronal emission may be delayed by a time interval that varies from event to event. The average of the temporal cross-correlation over all pixels smoothes out the peak.

The timing of individual events at different wavelengths is addressed here. Obviously, this is studied best in the strongest events. They constitute a small subset of events with thermal energies around $10^{26}$erg. By relating the coronal mass and heat inputs to emissions from layers below, the sequence of physical processes and the coupling of chromosphere and corona can be analyzed. Most of all, these properties may be compared to the well known association characteristics of full-sized flares in active regions.

The enhancement of the coronal emission measure in *regular flares* is generally interpreted as the result of impacting flare particles that transport primary flare energy from the corona into the chromosphere and heat it to millions of degree. The temperature of the evaporating plasma increases with the peak emission measure of the flare (Feldman et al. 1996). The relativistic tail of the primary electron distribution radiates gyro-synchrotron emission observable at microwave frequencies (e.g. review by Bastian et al. 1998). These electrons also excite O V and He I emissions (e.g. review Strong 1991).

However, the reconnection process that is generally assumed to trigger the flare energy release has never been observed in low-temperature lines and is generally believed to take place in the corona. What have blinkers and alleged low-temperature reconnection jets to do with coronal heating? Is coronal heating an effect of energy release at transition region temperatures or, inversely, are the transition region phenomena caused by energy release in the corona? We address these questions from the point of view of coronal heating, that means from the observed inputs into the corona, and by comparing them with associated emissions at other wavelengths.

## 2. Observations

Data from the Extreme ultraviolet Imaging Telescope (EIT) on board of the Solar and Heliospheric Observatory (SoHO) are used to measure the coronal emission measure. Simultaneous observations by the Very Large Array (VLA) at 3.6 and 6 cm were made, as well as by the Coronal Diagnostic Spectrometer (CDS) on SoHO in transition region lines. All instruments imaged an area of several arcminutes in size located in a quiet region in the center of the solar disk on July 12, 1996.

EIT is a normal-incidence, multi-layered mirror instrument (Delaboudinière et al. 1995). It imaged a 7'×7' area with a pixel size of 2.62"×2.62". The observing run lasted from 14:30 to 15:15 UT. EIT observed two wavelength bands, 171 Å and 195 Å, alternatively, resulting in a time resolution of 127.8 s at each line. The effective exposure time was 10 s for 171 Å and 20 s for 195 Å. The two bands include emission lines of Fe IX/X and Fe XII, respectively, with diagnostic capabilities for temperatures in the range of 1.1-1.9$\times10^6$K. Using the two bands, formal values for the coronal emission measure and temperature have been determined for each pixel. They have been evaluated for each observing time of Fe IX/X, interpolating the Fe XII flux for this time. There is one data gap at 14:53:35 UT. It may be noted here that in less than 1% of the pixels the line-ratio temperature is below 1.1$\times10^6$K and in none of them below 1.04$\times10^6$K. The emission measure, defined by the square of the electron density times the volume, is linearly proportional to the observed flux. The derived parameters are to be taken as formal values, representing weighted means over the sensitive temperature range.

The VLA has observed from 14:30 to 23:00 UT in D configuration. Only observations at the frequencies of 8.45 and 4.87 GHz are discussed here, to which we refer by wavelengths as 3.6 and 6 cm, respectively. The instrument observed at a single wavelength for 40 s with 10 s time resolution, and then changed to the other wavelength. The cycle time and effective resolution are 2.0 minutes. The wavelengths were observed alternating for 25 minutes, when the phase calibrations interrupted the observations for about 5 minutes. The field of view of the VLA is given by the single antenna (primary beam). It is circular and depends on wavelength. At 3.6 and 6.0 cm, the diameter of the field of view (FWHM) is 240" and 648", respectively. For long integration times the spatial resolution (FWHM) is roughly circular and for 3.6 and 6.0 cm it amounts to 7.1"and 12.5", respectively. The resolution decreases for snapshots. The radio data have been calibrated and cleaned as appropriate for solar observations.

CDS is a twin spectrometer applying normal and grazing incidence in the range 151 - 785 Å (Harrison et al. 1997). The He I and O V emission lines observed with normal incidence at 584Å and 629Å, respectively, were strong enough to be analyzed with sufficiently large signal to noise ratio. They originate from plasmas with temperatures around $2.0\times10^4$K and $2.4\times10^5$K, respectively. The instrument observed from 14:00 to 23:42 UT with a pixel size of 4.1"$\times$1.68" (East-West/North-South) in a field area of 41"$\times$240", using the 4.1"$\times$240" slit at 10 locations. The exposure time at each location was 20 s, resulting in a net time resolution of 4.4 minutes. For both EIT and CDS, linear interpolation has been used to correct for solar rotation. The spatial cross-correlation of images suggests that the final accuracy of the coalignment is better than $\pm$5". More information about the observations can be found in Benz & Krucker (1999).

The strongest increases in the coronal emission measure were selected for detailed analysis. As the overlap in time and field of view between all instruments is small, the comparisons EIT/VLA, and EIT/CDS were made separately, and the selected events are different.

The 23 most prominent enhancements of coronal emission measure with simultaneous radio coverage were selected from EIT data. As the CDS field of view is much smaller, it contained none of these events. Thus, the 6 brightest ones in the CDS field were selected for comparing coronal heating events with CDS data. The average thermal energy per event released in the first set is $1.30 \times 10^{26}$erg, in the second set it is only $6.8 \times 10^{25}$erg.

## 3. Results

The two sets of 23, resp. 6 largest events in emission measure have been integrated over their active area. The temperature and emission measure in the minimum preceding the event have been determined. The temperature and emission measure of the newly added material was then approximated after subtraction of the background contribution in each filter. The results are given in Tables I and II.

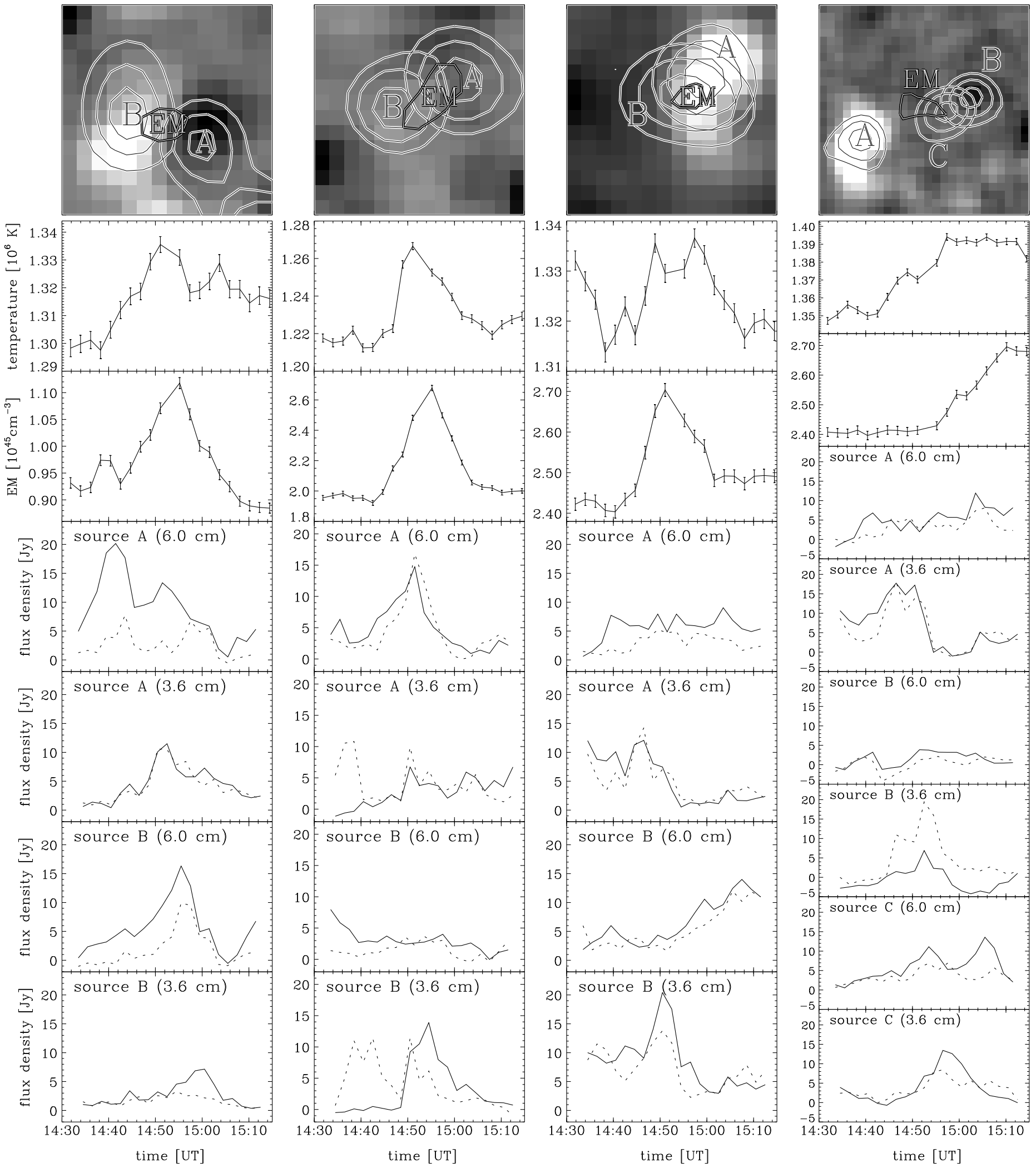

*Figure 1. Top row:* The pixels embraced by the solid curve were used for the calculation of the observed emission measure enhancements. The contours show the radio sources (labeled A, B, and C) at the 90%, 70% and 50% level of the peak in the image at maximum intensity and at the most intense frequency. The restoring beam size was chosen 14" FWHM in both frequencies. An MDI magnetogram with a pixel size of 2" and a 28"×28" field of view taken at 14:28:05 UT is underlain. *Second and third row:* Temperature and emission measure derived from EIT Fe IX and Fe XII lines. *Forth to ninth row:* Total radio flux density measured with the VLA for sources A, B, and C at right circular polarization (solid) and left circular polarization (dashed).

The time profiles of the emission measure integrated over the active area and the formal line-ratio temperature are shown in Fig. 1 for four examples. In all except one of the 29 events studied in detail, the Fe XII line peaked before the Fe IX/X lines or within the same 2 minute sampling time interval. Therefore, the derived temperature peaked at the time of the emission measure peak or before.

### 3.1. Comparison of Coronal Events with Radio Emission

Of the selected 23 emission measure enhancements, 22 had an associated radio source within a radius of 14" (10,000 km) and within 20 minutes in time. Some had more than one radio source associated and often, but not always, a source was seen at both 3.6 and 6 cm wavelength. A total of 35 associated radio events have been detected.

Table I gives an overview on the 23 events. Temperatures refer to the line-ratio values averaged over the whole event area. The background temperature and emission measure are determined at the time of the minimum in emission measure before the event. The event temperature $T_E$ is derived from the line ratio after subtracting the background values in the two lines. In one case marked with $*$ the temperature of the event is below one million degrees. Thus the derived value is questionable. The thermal energy of an event is

$$E_{th} = \frac{3}{2} n k_B \Delta T \approx 3 k_B T_E \sqrt{\delta\mathcal{M}} \sqrt{Ahq} \quad , \tag{1}$$

where $\delta\mathcal{M}$ is the observed emission measure increase, $A$ is the event area, $h$ the height and $q$ the filling factor. A product of $qh$=5000 km has been used for numerical estimates (incl. Tab.I), based on a well resolved event published by Benz & Krucker (1998). In the approximation of Eq. (1), $\Delta T \approx T_E$ has been assumed, implying that the heated material originates from the chromosphere. The inverse, the cooling of plasma above $2\times10^6$K, is unlikely as soft X-ray observations by Yohkoh have not found evidence for a sufficient amount of such plasma in the quiet corona (e.g. Benz et al. 1997). Note that the thermal energy does include neither a possible cooling by expansion, nor radiation and conduction losses in the rise phase. On the other hand, it assumes that the volume of the newly added material is small compared to the volume of the preexisting plasma. Thus Eq.(1) implies model assumptions.

Events 7, 1, 10, 11 (from left to right) are shown in Fig. 1. They all are associated with multiple radio sources occurring at different times and places. In all these cases the emission measure increase is located between two radio sources. Often, but not always the radio sources are related to magnetic elements in the photosphere. Discrepancies, such as the second and third event in Fig.1 may be due to the 20 minutes time difference between the start of these observations and the magnetogram.

Table I. Emission measure enhancements selected for comparison with radio emission. $N$ = event number, $A$ = active area in pixels (2.62"×2.62"), $n_{EM}$ = number of peaks in emission measure, $\tau$ = total duration in minutes, $T_{bg}$ = background temperature in units of $10^6$K, $T_E$ = temperature of new material in units of $10^6$K, $EM$ = emission measure enhancement in units of $10^{44}$cm$^{-3}$, $E_{th}$ = thermal energy of heating event in units of $10^{26}$erg, $n_r$ = number of radio sources, $\Delta t$ = delay of main Fe XII peak relative to radio peak.

| $N$ | $A$ | $n_{EM}$ | $\tau$ | $T_{bg}$ | $T_E$ | $EM$ | $E_{th}$ | $n_r$ | $\Delta t$ |
|---|---|---|---|---|---|---|---|---|---|
| | | | [min] | | | | [$10^{26}$erg] | | [min] |
| 1 | 10 | 1 | 21 | 1.22 | 1.41 | 4.4 | 1.6 | 2 | 3, 2 |
| 2 | 7 | 1 | 20 | 1.21 | 1.41 | 2.9 | 1.0 | 2 | 4, 0 |
| 3 | 51 | 3 | 40 | 1.21 | 1.32 | 5.8 | 3.9 | 2 | 8, -10 |
| 4 | 5 | 1 | 20 | 1.27 | 1.52 | 1.5 | 0.70 | 2 | 4, -1 |
| 5 | 7 | 1 | 25 | 1.27 | 1.39 | 1.9 | 0.94 | 1 | 4 |
| 6 | 9 | 1 | 11 | 1.22 | 1.63 | 2.7 | 1.4 | 2 | 2, 2 |
| 7 | 6 | 1 | 25 | 1.30 | 1.39 | 2.4 | 0.91 | 2 | 13, -1 |
| 8 | 4 | 1 | 15 | 1.38 | 1.49 | 1.1 | 0.45 | 2 | 11, 2 |
| 9 | 14 | 1 | 22 | 1.33 | 1.51 | 3.7 | 1.9 | 2 | 7, 0 |
| 10 | 15 | 1 | 16 | 1.32 | 1.56 | 2.9 | 1.8 | 2 | 6, 1 |
| 11 | 12 | 1 | 19 | 1.35 | 1.39 | 2.9 | 1.4 | 3 | 20, 17, 4 |
| 12 | 13 | 1 | 14 | 1.20 | 1.23 | 2.8 | 1.3 | 1 | 2 |
| 13 | 10 | 1 | 10 | 1.18 | 1.23 | 2.1 | 1.0 | 2 | -1, -6 |
| 14 | 9 | 1 | 23 | 1.22 | 1.33 | 1.7 | 0.93 | 1 | 11 |
| 15 | 13 | 2 | 21 | 1.31 | 1.34 | 2.9 | 1.4 | 1 | -1 |
| 16 | 6 | 1 | 12 | 1.21 | 1.08 | 1.6 | 0.65 | 1 | 3 |
| 17 | 10 | 2 | 27 | 1.28 | 1.32 | 3.2 | 1.3 | 0 | |
| 18 | 14 | 1 | 22 | 1.27 | 1.43 | 2.5 | 1.5 | 1 | 7 |
| 19 | 16 | 2 | 22 | 1.27 | 1.43 | 3.9 | 2.0 | 1 | 3 |
| 20 | 4 | 1 | 16 | 1.17 | 0.93* | 0.9 | 0.3* | 1 | 4 |
| 21 | 21 | 1 | 12 | 1.25 | 1.44 | 3.2 | 2.0 | 2 | 10, 3 |
| 22 | 7 | 2 | 22 | 1.20 | 1.33 | 1.6 | 0.78 | 1 | 2 |
| 23 | 9 | 1 | 15 | 1.35 | 1.12 | 1.7 | 0.80 | 1 | -1 |

The radio profiles have an effective rms noise of about 2 Jy. Thus, only peaks above 10 Jy are significant. Figure 1 demonstrates that the associated radio emission is rather complex.

### 3.1.1. *Spectrum*

Most radio sources are not spatially resolved. Since the restoring beamwidths are the same, the radio emission at the two wavelengths may readily be compared. Thermal emissions of the solar atmosphere at radio wavelengths do not decrease with frequency. Therefore, the sources labeled A in the first two events of Fig. 1 show a non-thermal behavior. Source B of the first event is also non-thermal, and source C of the fourth event is a mixture of peaks with positive and negative spectral slopes. In the following, spectra with *increasing* slopes are often referred to as being "thermal". Note however that this is a necessary, but not sufficient condition for thermal emission. In fact, the spectral maximum of some regular impulsive flares is below 3.6 cm, and in the observed increasing slope we may see the optically thick part of a gyro-synchrotron spectrum.

The numbers of thermal and non-thermal spectra are about equal: Of the 26 associated radio sources, for which a spectrum is available, 12 are consistent with thermal, 12 are non-thermal, and 2 change in time or are flat. This can be compared to a similar analysis at 3.6 and 2 cm wavelength by Krucker et al.(1997), who report 65% thermal, 25% flat, 5% non-thermal. The difference may be explained by the higher frequencies observed by those authors, favoring thermal emission, but more likely it is caused by their different selection based on the brightest radio events.

### 3.1.2. *Polarization and Mode*

The interferometric observation of polarization of weak events is limited by non-gaussian noise. A total of 27 events could be evaluated at their strongest frequency. We found 12 polarized beyond ±15%, and 9 were beyond ±30%. The significance of some extremely polarized radio emission, such as in the beginning of the second event in Fig. 1, is not clear.

The sense of polarization can be compared to the direction of the magnetic field. For a positive field and emission observed in the right hand circular mode, the radiation must have been emitted as extraordinary waves, assuming that the sense of polarization was not changed by a quasi-transverse region on its way. The same mode results from a negative field and left polarization, and ordinary mode from the other combinations.

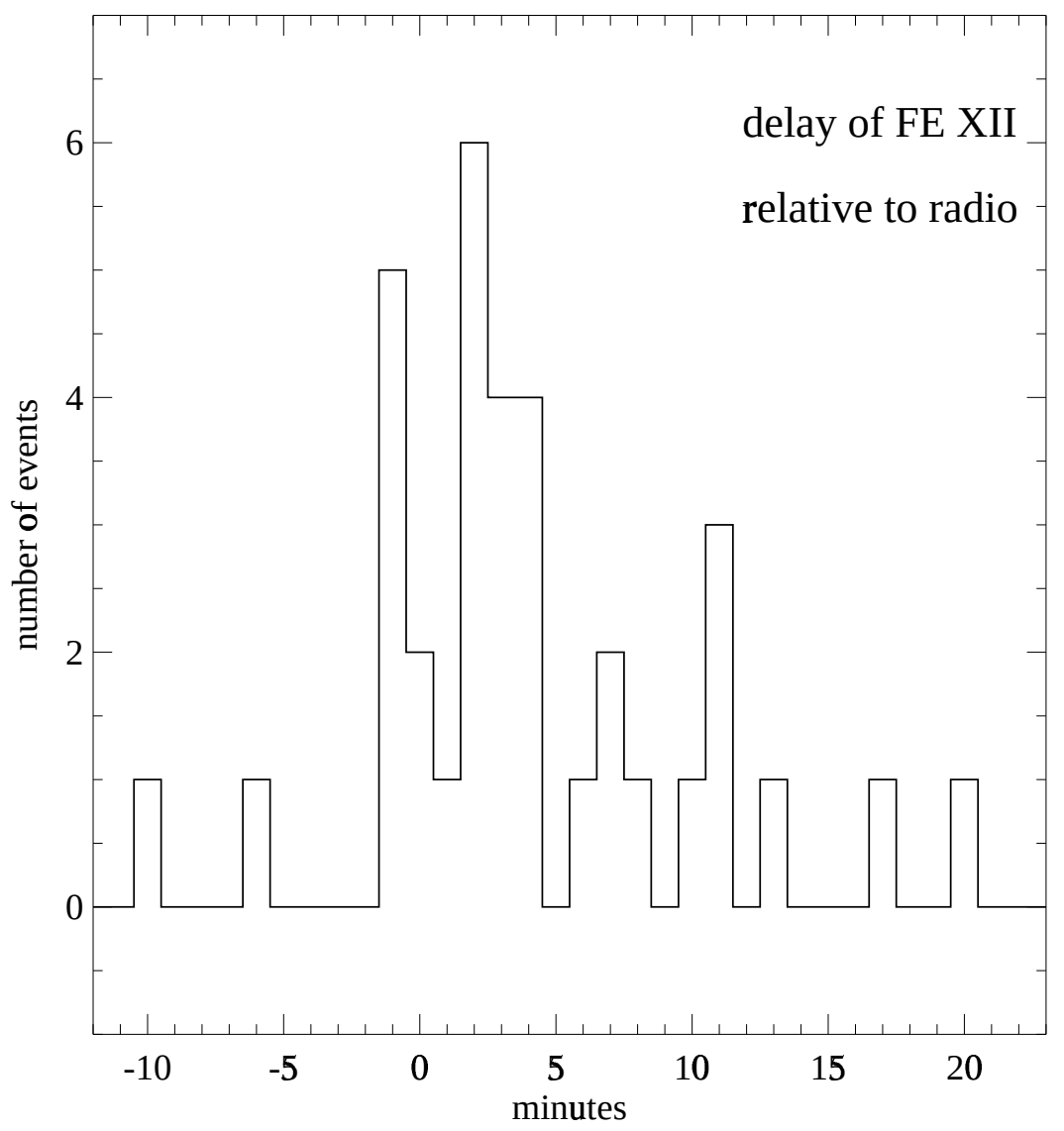

*Figure 2.* Distribution of the delays in the peak Fe XII photon count rate relative to the associated radio peak(s). The peak in the brighter of the two radio wavelengths is chosen for comparison. A positive delay means that the Fe XII emission peaks later than the radio emission.

The most reliable determinations of the magnetic polarity and the radio polarization are from the first and fourth event in Fig.1. Neglecting source C, we find extraordinary mode in 3 cases, ordinary mode in 1 case. Taking all cases except source C, the ratio is 5:3. Among the sources with non-thermal spectrum, the ratio is 3:1. Thus, the extraordinary mode is favored by these observations.

The frequent occurrence of spectra decreasing in frequency, the relatively high degree of circular polarization, and the prevalence of extraordinary mode are suggestive of the gyro-synchrotron process. There are definitely other cases as well, fully consistent with thermal radiation by free-free emission.

### 3.1.3. *Timing*

The timing information on the delay of the peak in Fe XII emission relative to radio waves in Tab. I is displayed in Fig.2. The timing uncertainty is half the sum of the time steps and, with some interpolation, approximately one minute. Out of 35 radio events, only 7 peaked later than Fe XII. The two events before -3 and after +15 minutes may be chance coincidences. The distribution is clearly asymmetric. The mean delays relative to thermal and non-thermal radio emissions are 3.3±4.0 and 3.9±3.9 minutes, respectively, neglecting the two most extreme delays on each side. Hence, the two distributions cannot be distinguished.

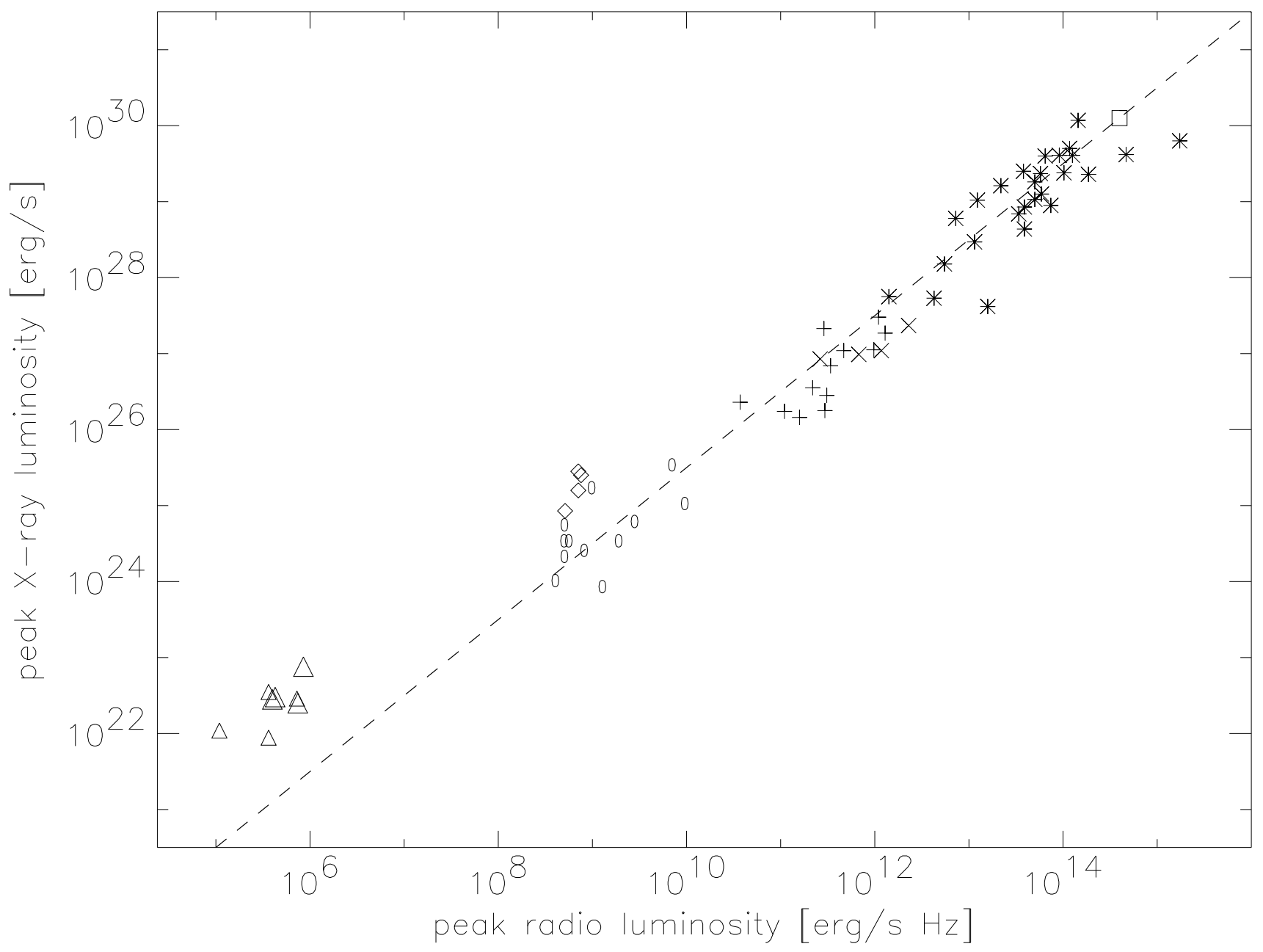

*Figure 3.* Comparison of the calculated peak soft X-ray and 3.6 cm radio luminosity of the heating events presented in Fig.1 (large △), heating events observed by Krucker et al.(1997, small △) with Yohkoh, microflares in active regions (◇, Benz et al. 1981; Fürst et al. 1982; 0, Gary et al. 1997), impulsive and gradual flares + and ×, Benz & Güdel 1994), quiescent dMe, dKe and By Dra stars (∗, Güdel et al. 1993), and a long duration flare on the dMe star EQ Peg (□, Kundu et al. 1988). The dashed line is a fitting curve with slope 1 to the dMe stars given by Eq.(2).

The wide distribution of delays is clearly influential in reducing the cross-correlation peaks between radio and Fe XII and may help to explain the low values reported by Benz & Krucker (1999).

### 3.1.4. *Total Coronal Luminosity*

Benz & Güdel (1994) have found a general relation between soft X-ray luminosity, $L_x$ in erg s$^{-1}$, and gyro-synchrotron radio luminosity, $L_r$ in erg s$^{-1}$ Hz$^{-1}$, from the smallest flares in active regions (called microflares by those authors), regular impulsive flares, gradual flares up to huge stellar flares and "quiescent" emissions of dMe and dKe stars, and BY Dra-type binaries. The relation is

$$L_x/L_r \approx 10^{15.5\pm0.5} \quad [\mathrm{Hz}] \ . \qquad (2)$$

The reason for the proportionality of the radio and X-ray emissions seems to be that the same physical processes of flare energy release, particle acceleration, and dissipation occur over a large range of parameters that span ten orders of magnitude including also RS CVn, FK Com and T Tau stars (not shown in Fig. 3). Do the heating events of the quiet corona comply with this relation?

Given the emission measure and temperature of a heating event (cf. Tab.I), its peak soft X-ray luminosity has been estimated using the SPEX software package (Kaastra et al. 1996). Solar abundances and ionization equilibrium are assumed. The following approximation to the SPEX curve valid for temperatures around 1.5 MK was used:

$$L_x \approx \ 1.072 \times 10^{-19} \delta\mathcal{M} \ T_E^{-0.4675} \quad [\mathrm{erg/s}] \ . \qquad (3)$$

We have adjusted the luminosity to the convention of ROSAT so that the heating events can be compared to the data of Benz & Güdel (1994). The operation requires a multiplication of Eq.(3) by 0.72.

The radio flux density of all coincident sources at 3.6 cm has been added in time, and its peak value was transformed to spectral luminosity. This value is an upper limit to the gyro-synchrotron emission. The wavelength corresponds to the one used by Benz & Güdel (1994).

The four heating events presented in Fig.1 and some events observed in soft X-rays (Krucker et al. 1997) are shown in Fig.3 together with X-ray and radio luminosities observed in large solar and stellar flares. Note that the X-ray luminosities derived here from EUV observations and the actually observed ones from Krucker et al.(1997) are consistent. The heating events are "radio-poor" relative to relation (2) by more than an order of magnitude. A slight deviation to less radio emission may be apparent already in microflares of active regions. The peak $L_x/L_r$ relation of heating events clearly deviates from the scaling observed for large flares.

Table II. Emission measure enhancements selected for comparison with transition region line emissions. The source area refers to the active area in EIT pixels (2.62"×2.62").

| source nr. | area [pix.] | total dur. [min] | temp. event [$10^6$K] | EM increase [$10^{44}$cm$^{-3}$] | thermal energy [$10^{26}$erg] | delay of Fe XII O V [min] | He I [min] |
|---|---|---|---|---|---|---|---|
| 1 | 8 | 17 | 1.30 | 0.9 | 0.6 | 4 | 4 |
| 2 | 11 | >16 | 1.30 | 1.0 | 0.8 | 5 | 4 |
| 3 | 15 | 12 | 1.21 | 1.9 | 1.1 | 7 | -3 |
| 4 | 8 | 16 | 1.39 | 1.4 | 0.8 | 5 | 9, -1 |
| 5 | 2 | 15 | 1.36 | 0.4 | 0.2 | 8 | 5 |
| 6 | 9 | 12 | 1.19 | 0.8 | 0.6 | -3 | 2 |

### 3.2. Comparison of Coronal Events with Line Emissions from the Transition Region

The six largest enhancements in emission measure observed in the common field of view of EIT and CDS are listed in Table II. The selection was made from EIT events, and their active area was searched for a CDS event. All apparently associated CDS pixels have then been selected for analysis. As the CDS pixels are larger and the sources often slightly displaced, the source areas of EIT and CDS do not completely overlap. The average pixel area and total duration are generally smaller than in the previous set, and so is the thermal energy input into the corona. The delay of the Fe XII peak relative to O V and He I is accurate to within half the sum of the two time steps, or $\pm$ 3 minutes. We have improved the accuracy somewhat by visual interpolation.

Four of the 6 events of this second set have associated radio peaks preceding the emission measure peak; one radio peak is delayed, one is questionable. All radio peaks of the second set are weak, however, and thus are not presented here.

Figure 4 displays the six events with CDS coverage. The EIT measurements of the source properties in area, duration, temperature, and coronal emission measure are given in Tab. II. The time of the coronal Fe XII peak relative to the peaks at transition region lines as observed by CDS are listed. Note in particular:

1. The changes in the coronal emission measure are far less spectacular than in the transition region lines. The former changes by at most 20%, but O V varies by more than factors of 2.

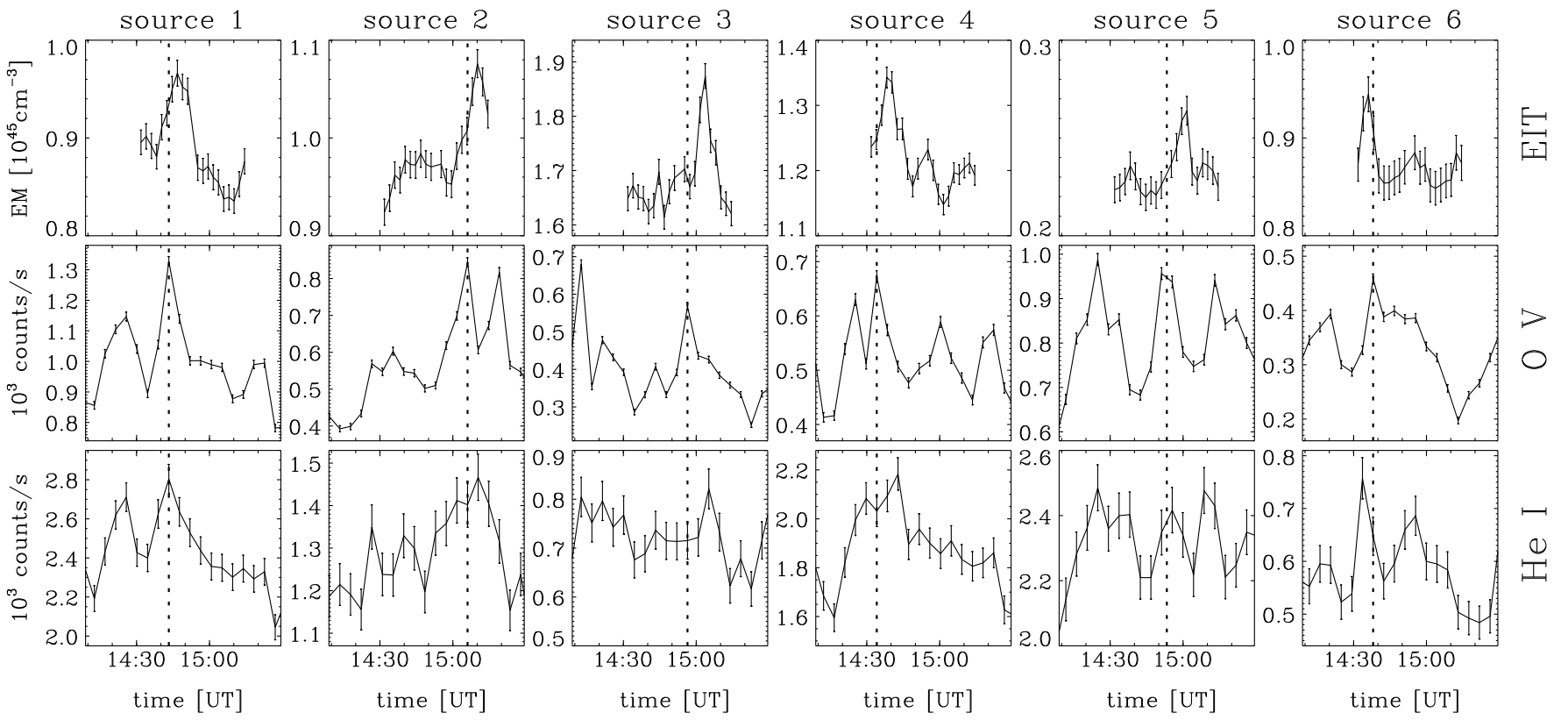

*Figure 4.* *Top row:* Emission measure of heating events vs. time. The active pixels have been summed and EIT measurements of Fe IX/X and Fe XII were used. *Second and third row:* Count rates in O V and He I lines observed with CDS from the same area.

2. The number of transition region brightenings is larger than the number of coronal enhancements.
3. Due to the large number of brightenings of the transition region, the significance of their association with coronal events is not always clear.
4. O V is best associated with the coronal events. It precedes the coronal emissions in 5 of 6 cases. The average delay of emission measure increase is 4.2 minutes in agreement with the value found by cross-correlation of all pixels (Benz & Krucker 1999). The individual peaks, however, scatter from 8 to -3 minutes.
5. He I precedes the coronal emission measure in 4 out of the 6 cases and by an average value of 1.8 minutes.

### 3.3. Line Shape During Heating Events

The analysis of O V line profiles at 629 Å often reveal changes in line shape. Figure 5 displays difference profiles of the O V line in velocity (or wavelength). Negative velocities denote blue shift. The line profile averaged over the same area and the total observing time has been subtracted to enhance the visibility of the changes. As the line is strongest at zero shift, the counts there are largest and so are the error bars in Fig.5. The subtraction of the background line profile such as presented in Fig.5 is unconventional. The involved mathematics is outlined in the appendix.

Table III. Line shifts in O V of Source 3.

| time UT | separation Δ of peaks [km/s] | line shift |
|---|---|---|
| 14:43:17 | | no peaks |
| 14:43:38 | 183 | blue |
| 14:52:01 | 278 | blue |
| 14:56:24 | 167 | blue |
| 15:00:48 | 139 | blue |
| 15:05:11 | 222 | red |
| 15:09:33 | 193 | red |
| 15:13:56 | 306 | red |

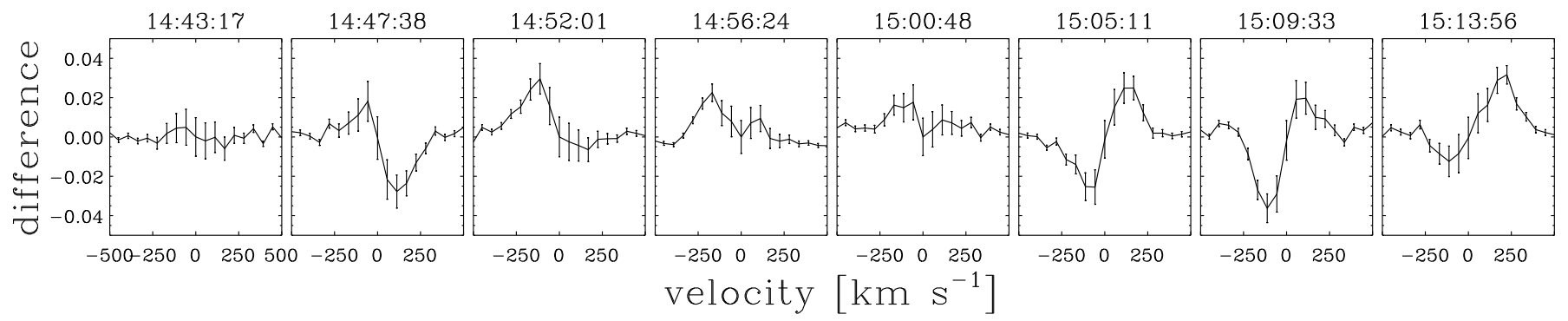

*Figure 5.* Average line profile of O V over the active area of Source 3 as observed by CDS. The difference relative to the average profile before the event is shown. The velocity is calculated from the observed wavelength. The starting time of each raster given at the top of each plot increases from left to right.

The series of 8 line profiles in Fig. 5 covers the rise and fall of a heating event over a total time of 30 minutes. The separation of positive and negative peaks in Fig. 5 indicates the spectral shift of the line. The first plot of the background before the event shows a nearly flat profile. The second one originates from a pronounced, blue-shifted line. The blue shift becomes less clear and turns into a pronounced red shift in the decaying part of the event. Note that the observed general line shift reverses sign at the peak of the emission measure enhancement at 15:00 UT. As derived in the appendix, only separations beyond about 200 km/s can be considered to reflect true Doppler motions. Values in Tab. III below this are probably caused by variations of the line profile during the event. The spectral resolution is 50 km/s, and with interpolation the nominal error in Tab. III is smaller than ±17 km/s.

The line profiles of the other sources are less systematic, and the subtraction analysis did not always yield a simple picture. Nevertheless, three of the six selected events suggest a blue shift in the rise phase and a red shift in the decay phase of a heating event, two events showed only red shifts and one only blue shift. The shifts reach velocities of 200-280 km/s in the rise phase and -310 km/s in the decay phase in the best event, presented in Fig.5.

## 4. Discussion

In regular flares, the thermal flare plasma reaches temperatures beyond $10^7$K and thus much higher values than the background corona. Feldman et al. (1996) have found the peak thermal flare temperature to increase with the emission measure. The heating events in the quiet corona studied here have a small emission measure and do not much exceed the background temperature of about $1.3\times10^6$K. Thus they concur with the trend in flares. As pointed out already by Krucker et al. (1997) heating events are, however, not consistent with the exponential relation of Feldman et al. extrapolated to smaller events.

We find a generally good association of coronal emission measure enhancements with brightenings of emissions from lower layers. The amplitude of the associated emission varies considerably in the analyzed events, showing more variation in the transition region emissions. This may be a reason for the relatively bad association of the inverse relation, i.e. of the association of transition region brightenings with coronal peaks.

We furthermore note a large spread of delays in the timing of peak values of the different emissions. In the following, the parameters observed in heating events are compared with regular, full-sized flares from active regions that are typically four orders of magnitude more powerful.

### 4.1. Coronal Emission Measure and Centimeter Radio Emission

The radio emission of flares has been compared to the emission measure of the thermal flare plasma by Neupert (1968), who used observations at centimeter wavelength similar to these observations. Flare centimeter waves are preferentially emitted by the gyro-synchrotron mechanism. Neupert found a delay of the soft X-ray peak by 0.5 to 5 minutes relative to the radio peak. More recently, soft X-rays or coronal lines have been compared to hard X-rays. Hard X-rays (>20 keV) and centimeter radio emission peak usually within a few seconds (e.g. Takakura et al. 1983). Dennis & Zarro (1993) find delays of soft X-ray peaks by 2.5 to 8 minutes relative to the hard X-ray peak in impulsive flares and more than 40 minutes in gradual flares. Benz & Güdel (1994) report an example of an impulsive flare with a time delay of 3 minutes of the thermal flare emission measure relative to centimeter emission, and an example of a gradual flare with a delay of 66 minutes.

The observed range of delays of the Fe XII peak relative to the associated radio peak of heating events is -1 to about 12 minutes (Fig.2) and thus overlaps only partially with the observed range of delays seen in flares. The peak around zero delay is different and may indicate the presence of a different mechanism causing thermal emission.

The radio emission of flares at centimeter waves is completely dominated by non-thermal gyro-synchrotron radiation. In heating events on the other hand, the non-thermal spectra are as frequent as those that are consistent with thermal emission. This is not surprising since the background radio emission of the quiet Sun in centimeter waves consists of thermal emission. A 32% enhancement in density of the transition region at constant temperature or a 47% enhancement in temperature at constant density produces a brightening by 23% according to model calculations by Benz et al. (1997).

### 4.2. Coronal Emission Measure and Transition Region Line Emission

The transition region line emission, in particular the O V line, has been reported to peak within seconds of the centimeter radio emission in full-sized flares (Castelli & Richards 1971, and many others), and thus precedes the coronal emission measure increase by several minutes. More recently, lines from the transition region during flares have been compared with hard X-rays. Poland et al. (1982) and Orwig & Woodgate (1986) find excellent agreement between O V line emission and hard X-rays within less than one second. It suggests that the two emissions are produced close to each other, namely in the upper chromosphere or transition region, presumably as a result of an impinging particle beam.

### 4.3. Line Shape of Transition Region Lines in Flares

Mass motions have been extensively studied in high-temperature lines of regular flares. A large number of investigations on energy transport has been carried out using observations from P78-1, SMM, and Hinotori (review e.g. by Zarro 1992). Most of the work has been done on the Ca XIX and Fe XXV lines of the relatively hot thermal flare plasma.

The observations indicate nonthermal broadening and asymmetry by blue shifts. The blue asymmetry is most pronounced during the rise phase of the soft X-ray emission and can be modeled by an emission component that is Doppler shifted by an upward velocity of 300-400 km/s (Antonucci et al. 1984). The Ca XIX line is also broadened, presumably caused by unresolved elements of motion moving at different velocities. The broadening reaches a maximum of $\gtrsim$ 200 km/s at the peak of the hard X-ray emission as compared to the preflare level of $\lesssim$ 50 km/s in active regions (Antonucci & Dennis 1983).

Much less is known about mass motions in transition region lines during flares. Bruner & Lites (1979) measured the shift of the C IV line at 1548 Å and found downflows of 30 km/s. Brekke et al. (1996) report downflows in C IV and Si IV during the peak of centimeter radio emission of 50 km/s.

The observed line width in O V of $\pm$140 km/s far exceeds the thermal ion velocity at the temperature of maximum formation, 11.1 km/s, and seems to indicate a high level of turbulence. Observing active region loops, Brekke et al.(1996) have noticed even larger line widths of $\pm$170 km/s .

The analysis of the O V line is preliminary. The line shifts in relation with coronal heating events will be further investigated in other transition region lines and with higher spectral resolution.

## 5. Conclusions

The trend in individual heating events analyzed here suggests that the O V transition region brightenings occur minutes before the radio emission and the cooler He I line emission. The radio and He I radiations seem to brighten before the coronal Fe IX/X and Fe XII emissions. The range of individual delays is several minutes. The mean values generally agree with the values derived by temporal cross-correlation of common pixels (Benz & Krucker 1999).

The radio emission associated with heating events may be thermal in up to half of the cases studied here. Nevertheless, some of the heating events are associated with centimeter waves that show the typical characteristics of gyro-synchrotron emission such as a spectrum increasing with wavelength, relatively high circular polarization, and a peak that precedes the thermal emission. This radio emission represents strong evidence for the presence of mildly relativistic electrons in heating events.

However, the gyro-synchrotron emission associated with heating events differs from flare centimeter emission at least quantitatively. The ratio of peak centimeter emission (including both gyro-synchrotron and potentially thermal emissions) to peak soft X-ray luminosity is more than an order of magnitude below regular flares (Fig. 3). This low intensity of gyro-synchrotron emission and the different radio background in the heating events compared to active region flares seem to make it possible to detect associated thermal events in quiet regions.

Finally, the line profiles in O V show (blue-shifted) upflows in the rise phase of half of the studied heating events and downflows in the decay phase. The velocities exceed 300 km/s. These observations are suggestive of chromospheric evaporation and coronal cooling, although no mass motion in transition region lines of this magnitude and in the observed time sequence has ever been reported for flares.

In conclusion, heating events hold many properties in common with flares, but the timing and flux ratios to emissions from layers below may vary in a large range, just as in flares or even larger. Some of the differences between heating events and flares can be explained by the different background emissions in quiet and active regions. This work has not uncovered any principal differences between flares and heating events except for size and rate of occurrence.

5.0.1. *Acknowledgements*

SoHO is a joint project between the European Space Agency (ESA) and NASA. We thank the EIT team for their excellent work and in particular J.-P. Delaboudinière for clarifying discussions on the EIT instrument, and T. S. Bastian, B.J. Thompson and D. Pike for help with the observations. We thank A. Pauluhn and the referee for a careful and critical reading of the manuscript. EIT was funded by CNES, NASA, and the Belgian SPPS. The Very Large Array is operated by Associated Universities, Inc. under contract with the US National Science Foundation. We acknowledge the use of CDS and MDI data, both from the SoHO satellite. The work at ETH Zurich is financially supported by the Swiss National Science Foundation (grant No. 20-53664.98).

## 6. Appendix

Figure 5 displays the difference spectra during a heating event and the preexisting background spectrum. The subtraction of the two spectral line profiles needs some mathematical comments. In the following the line profiles are assumed to be Gaussian and of the form

$$L(\nu) \ = \ L_0 \exp\left[-\left(\frac{\nu-\nu_0+\sigma}{\Gamma}\right)^2\right] \quad , \tag{4}$$

where $\Gamma = 0.5(\ln 2)^{-1/2}\nu_{FWHM}$ is assumed to be constant during the event, $\nu_{FWHM}$ is the full width at half maximum of the line, $\nu_0$ is the peak frequency of the unshifted line, and $\sigma$ is the line shift due to Doppler motion.

The subtraction of two lines with different shifts $\sigma_1$ and $\sigma_2$ leads to peaks at the frequencies that solve the transcendental equation

$$\frac{\bar{\nu} - \Sigma/2}{\bar{\nu} + \Sigma/2} \exp\left[-(\bar{\nu} - \Sigma/2)^2 + (\bar{\nu} + \Sigma/2)^2\right] \; = \; 1 \quad , \tag{5}$$

where we have defined

$$\bar{\nu} \; = \; \frac{\nu - \nu_0 + \frac{1}{2}(\sigma_1 + \sigma_2)}{\Gamma} \quad , \tag{6}$$

$$\Sigma \; = \; \frac{(\sigma_1 - \sigma_2)}{\Gamma} \quad . \tag{7}$$

Eq.(5) has two solutions near $\nu_0$ corresponding to a positive and a negative peak. They are separated by a frequency interval $\Delta$, having the asymptotic values of

$$\Delta \approx \Sigma \quad \text{for} \;\; \Sigma^2 \gg \Gamma^2 \tag{8}$$

$$\Delta \approx \sqrt{2}\Gamma \quad \text{for} \;\; \Sigma \lesssim \Gamma \tag{9}$$

Note that the separation does not vanish as the difference in drift approaches zero, because the largest differences of slightly shifted Gaussian profiles occur at the steepest slopes. The observed separations $\Delta$ of spectral peaks in Fig.5 are listed in Tab. III.

A $\Gamma$ of 140 km/s has been measured. Assuming constant Gaussian line profiles and applying the above mathematical relations, the smallest value for $\Delta$ should be $\sqrt{2}\Gamma$. Only separations beyond about 200 km/s can be considered to reflect true Doppler motions. Values in Tab. III below this are probably caused by variations of the line profile during the event.